\documentclass[12pt]{article}
\usepackage{latexsym,epsfig}
\usepackage{graphicx}
\usepackage{hyperref}
\usepackage{polylib}
\usepackage{multirow}

\newcommand{\abs}[1]{\left|#1\right|}
\newcommand{\set}[1]{\left\lbrace#1\right\rbrace}
\newcommand{\ceil}[1]{\left\lceil#1\right\rceil}

\newcommand{\comment}[1]{\relax}

\graphicspath{{./graphics/}}
\polypath{./graphics/}

\usepackage[a4paper,top=20mm,bottom=10mm,left=15mm,right=15mm]{geometry}
\usepackage[T1]{fontenc}
\usepackage{newtxtext,newtxmath}
\usepackage[scaled=0.92]{helvet}
\usepackage{microtype}
\usepackage{hyperref}
\usepackage{fancyhdr}
\usepackage{titlesec}

\hypersetup{colorlinks=false,pdfborder={0 0 0}}
\titleformat{\section}
  {\sffamily\bfseries\fontsize{12}{14}\selectfont}
  {\thesection}{0.45em}{}
\titlespacing*{\section}{0pt}{18pt}{9pt}

\newcommand{\papertitle}{Counting Polyominoes, Revisited: \,
            Corrigendum and Addendum}
\newcommand{\authors}{Gill Barequet \,\(\cdot\)\, Gil Ben-Shachar}

\begin{document}
\thispagestyle{fancy}

\vspace*{-13mm}
\noindent\rule{\textwidth}{0.7pt}

\vspace{4mm}
{\sffamily\bfseries\fontsize{17}{20}\selectfont
\papertitle\par}

\vspace{4mm}
{\sffamily\bfseries\fontsize{12.5}{15}\selectfont
\authors\par}

\vspace{1mm}
{\fontsize{9.25}{10.5}\selectfont
Dept.\ of Computer Science, The Technion---Israel Institute of Technology,
Haifa 3200003, Israel, \{barequet,gilbe\}@cs.technion.ac.il
\par}

\vspace{5mm}
{\sffamily\bfseries\fontsize{11.5}{15}\selectfont
Corrigendum: \, \boldmath{$A(67)$}
\par}

\vspace{3mm}
{\fontsize{10}{12.2}\selectfont
In the original paper, we extended the enumeration of polyominoes from size~56 to size~70.  There was a small typo in the value for size~67, as follows: \\
\begin{tabular}{ll}
Published: & 288,146,737,936,083,547,752,609,866,746,184,117,54\textbf{6}; \\
Correct:   & 288,146,737,936,083,547,752,609,866,746,184,117,54\textbf{4}.
\end{tabular} \\
That is, the published value was too large by~2.

\medskip

The error was kindly brought to our attention by Christopher Honey
(email: \texttt{chris.honey@jhu.edu}), who noticed (with some help of
Claude Fable) that the published value did not match the congruence modulo~4 (which should have been~0 rather than~2) predicted by a combinatorial formula
relating the count of all polyominoes of size~67 to the counts of polyominoes
of size~67 in the various symmetry classes.

\medskip

Upon inspecting the code, which implemented a ``45$^\circ$-rotated'' version of
Jensen's algorithm~[3], we found no error or bug.  Moreover, the
output of our program agreed with the first~56 values computed by Jensen at the
beginning of the 2000s.
Nevertheless, we were able to identify the source of the erroneous value.  The
algorithm counted polyominoes in a variety of bounding boxes, and all the
intermediate counts appear to be correct.  However, either their sum was computed
incorrectly, or it was computed correctly but was subsequently transcribed
incorrectly into the manuscript.
\par}


\vspace{5mm}
{\sffamily\bfseries\fontsize{11.5}{15}\selectfont
Addendum: \, Complexity Analysis
\par}

\vspace{3mm}
{\fontsize{10}{12.2}\selectfont
The dominant (exponential) factor in the running time and in the memory consumption of Jensen's algorithm is the number of signatures that appear during the course of the
algorithm.
As mentioned by Barequet and Moffie~[1], the presence of kinks
contributes only a constant factor, hence, we analyze here only boundaries without kinks.

\medskip

Unlike in the original version of the algorithm (on the non-rotated lattice), consecutive
occupied cells on the boundary do not necessarily belong to the same component.
Hence, we are looking for the number~$T(n)$ of strings of length~$n$ over an alphabet of
size~$n{+}1$, consisting of~$n$ letters (identifiers of connected components) and a
special symbol~0 (for empty cells), such that, after deleting all 0s, the remaining
string contains no crossing subsequence pattern of the form $abab$, with distinct
letters $a \neq b$.
Strings are identical up to relabeling of the letters except the special letter 0.
Consider strings that have~$k$ letter-positions and $n{-}k$ 0s, and that do not contain a
crossing pattern~$abab$.  The number of such strings (for all possible values of~$k$) is
\[
	T(n) = \sum_{k=0}^n \Biggl( \binom{n}{k} \frac{1}{k+1}\binom{2k}{k} \Biggl),
\]
where~$\binom{n}{k}$ is the number of ways in which we can choose the letter positions,
and $\frac{1}{k+1}\binom{2k}{k}$ (the $k$th Catalan number) is the number of length-$k$
strings that do not contain any crossing pattern.

\medskip

We now find the asymptotic behavior of~$T(n)$.
Let $C(z)=\sum_{k \geq 0} C_k z^k = \frac{1-\sqrt{1-4z}}{2z}$ be the ordinary generating
function of the Catalan numbers.  Since
\(
    T(n) = \sum_{k=0}^n \Big( \binom{n}{k} C_k \Big)
\)
is the (unweighted) binomial transform of $C_k$, its ordinary generating function is
\[
    F(x) = \sum_{n \geq 0} T(n)x^n = \frac{1}{1-x} \, C\!\left(\frac{x}{1-x}\right) =
        \frac{1}{2x}\left(1-\sqrt{\frac{1-5x}{1-x}}\right).
\]
The dominant square-root singularity at $x=\frac{1}{5}$ then implies, by the transfer
theorem~[2, Ch.~VI, Thm.~VI.1], that
\[
    T(n) \sim \frac{5\sqrt{5}}{8\sqrt{\pi}} \cdot \frac{5^n}{n^{3/2}}.
\]
Since $T(n)$ is an upper bound on the number of signatures,
the dominant exponential term in the latter number is at most~$5^n$.


\vspace{5mm}
{\sffamily\bfseries\fontsize{11.5}{15}\selectfont
References
\par}

\vspace{3mm}
{\fontsize{10.0}{10.5}\selectfont
\,\,\,[1]\,\,\,{\sc G. Barequet and M. Moffie},
On the complexity of Jensen's algorithm for counting fixed polyominoes.
\emph{J. of Discrete \mbox{~~~~~~~~~}Algorithms},
5~(2), 348--355, June 2007.

\,\,\,[2]\,\,\,{\sc F. Flajolet and R. Sedgewick},
\emph{Analytic combinatorics},
Cambridge University Press,
2009.

\,\,\,[3]\,\,\,{\sc I. Jensen},
Counting polyominoes: A parallel implementation for cluster computing,
{\em Proc.\ Int.\ Conf.\ on Computational \mbox{~~~~~~~~}\,Science},
Melbourne, Australia,
{\em LNCS}, 2659,
Springer-Verlag, Heidelberg, 203--212, August 2003.
\par}

\comment{
{\fontsize{10.0}{10.5}\selectfont

}
}

\vspace{2mm}
{\fontsize{8.8}{10.5}\selectfont
\textbf{Author’s note.}
This is an author-prepared corrigendum and addendum to the paper cited above. It has not
been published by Algorithmica and has not undergone the journal’s usual submission,
review, and acceptance process.
\par}

\newpage


\fancyhead[L]{}

\title{
   \vspace{-5mm}
   \fbox{\large This paper was published in Algorithmica 88,
                Article 51 (2026).} \\[10mm]
   Counting Polyominoes, Revisited\thanks{%
      Work on this paper has been supported
      in part by BSF Grant~2017684.
      A preliminary version of this paper~\cite{barequet2024counting}
      appeared in ALENEX'24.
   }
}

\author{
   Gill Barequet\thanks{
      Dept.\ of Computer Science,
      Technion---Israel Institute of Technology,
      Haifa~3200003, Israel.
      E-mail: \texttt{\{barequet,gilbe\}@cs.technion.ac.il}
   }
\and
   G. Ben-Shachar$^\dag$
}
\date{}

\maketitle


\begin{abstract}

   A \emph{polyomino} is an edge-connected set of squares on the square
   lattice.
   In this paper, we improve the Conway-Jensen polyomino-counting algorithm
   by considering bounding boxes on the square lattice rotated
   by~$45^\circ$ instead of on the regular unrotated lattice.
   This allows us to extend significantly the count of polyominoes from~56
   to~70 terms.

   \bigskip

   
\end{abstract}


\section{Introduction}

Polyominoes of size $n$ are edge-connected sets of $n$ cells on the square
lattice.  Methods for counting them have been studied for more than~60
years in statistical physics in the context of percolation
processes~\cite{broadbent1957percolation}, and this problem has since then
been an active research topic in enumerative combinatorics.
Counting polyominoes was stated as Problem~37 in The Open Problems Project
(see \texttt{https://topp.openproblem.net/p37}).
In the mathematics literature, the notation $A(n)$ (Sequence~A001168 in the
On-line Encyclopedia of Integer Sequences~\cite{sloane2018line}) is commonly
used for the number of \emph{fixed} polyominoes, that is, polyominoes that
are distinct up to translations.

More than half a century ago, Klarner~\cite{klarner1967cell} proved that
the limit $\lambda = \lim_{n \to \infty} \sqrt[n]{A(n)}$ exists, henceforth
coining~$\lambda$, the growth constant of~$(A(n))$, as ``Klarner's
constant.''
The convergence of $A(n+1) / A(n)$ to $\lambda$, as $n$ tends to
infinity, was proved only three decades later by Madras~\cite{madras1999pattern}.
It is widely believed that~$\lambda \approx 4.06$, and the currently best
estimate, $\lambda = 4.0625696 \pm 0.0000005$, is due to
Jensen~\cite{jensen2003counting}.
The best-known lower and upper bounds on~$\lambda$ are
4.0025~\cite{barequet2016lambda} and
4.5252~\cite{barequet2021improved}, respectively.

Early algorithms counted polyominoes by generating them systematically,
however, with repetitions, hence most of their running time was spent on
identifying and eliminating duplicates.
The first efficient algorithm for counting polyominoes was presented
in~1981 by Redelmeier~\cite{redelmeier1981counting}.  
His algorithm, which was based on a method for enumerating connected
subgraphs that contain a marked vertex in a graph, counted polyominoes by
generating them one by one \emph{without repetitions}.
It was then used for counting polyominoes up to size~$n=24$.

The second-generation algorithm, suggested by
Jensen~\cite{jensen2001enumerations,jensen2003counting} (see also
Ref.~\cite[Appendix]{jacobsen2009polygons}), provided
a listing of of~$A(n)$ up to $n=56$.
As Jensen indicates~\cite[p.~205]{jensen2003counting}, the algorithm he
used was based on the finite-lattice method used by
Conway~\cite{conway1995enumerating} for the calculation of series expansions
for percolation problems, and was similar to methods devised by
Enting~\cite{enting1980generating} for enumeration of self-avoiding polygons.
Therefore, we hereafter refer to this algorithm as the
``Conway-Jensen algorithm.''
The genuine contribution of the algorithm was in counting
polyominoes \emph{without} actually producing them.
In this paper, we present a significant improvement to this algorithm,
and extend considerably the known count of polyominoes up to size~70.

The paper is organized as follows.
In Section~\ref{sec:jc-algorithm}, we briefly describe the Conway-Jensen
algorithm.
Then, in Section~\ref{sec:our-improvement}, we describe our improvement
to the algorithm, and detail further in Section~\ref{sec:implementation}
implementation details and differences between the two versions of the
algorithm.
In Section~\ref{sec:comparison}, we proceed to a comparison between the
two versions.
In Section~\ref{sec:results}, we provide the new counts of polyominoes.
We end in Section~\ref{sec:conclusion} with some concluding remarks.


\section{The Conway-Jensen Algorithm}
\label{sec:jc-algorithm}

We provide here the main ingredients of the Conway-Jensen algorithm.
The interested reader is referred to
Refs.~\cite{jensen2001enumerations,jensen2003counting} for more details.

The algorithm is based on a transfer-matrix method which is efficient
mainly because it does not produce all polyominoes.
Polyominoes are built imaginarily, column by column, from left to right,
and in every column, squares are considered, one by one, from top to bottom.
Instead of keeping in memory all possible polyominoes, the algorithm
maintains all possible \emph{right boundaries} of polyominoes
partially built within axis-aligned boxes.
(Here, ``partially built'' means that during the course of the algorithm,
the constructed shapes are possibly made of yet disconnected components.)
Figure~\ref{fig:boundary}
\begin{figure}
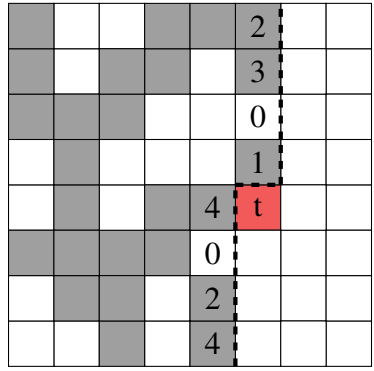

   \centering
   \drawpoly[scale=1.5]{partial_orig.txt}
   \caption{A partly-built 25-cell polyomino with signature
            ``\texttt{23014024}''.
            The current boundary is marked with a dashed line and the next
            cell to be considered, $t$, is colored in red.}
   \label{fig:boundary}
\end{figure}
shows the right boundary of a sample partially-built polyomino.

The algorithm keeps, for every possible right boundary, counts of
all options for the connectivity (that is, the existence of paths of
4-neighboring cells) between cells of the boundary through cells found to
the left of it.
The composition of a boundary and its connectivity information is called a
``signature,'' which is encoded by a 5-letter alphabet:
`0' encodes an empty cell;
`1' encodes a singleton boundary cell;
and if at least two boundary cells belong to the same component, then
`2' encodes the highest,
`4' encodes the lowest,
and `3' encodes all the other boundary cells (if exist) that belong to
this component.\footnote{
   In his implementation of the Conway-Jensen algorithm,
   Knuth~\cite{knuth2001polynum}
   used the alphabet~``01(-)'' instead of~``01234''.
}
The signature encodes only the boundary configuration and connectivity
information.  The position of the kink (the currently considered cell)
is represented implicitly by the state of the algorithm rather than being
encoded in the signature itself.  When the current cell is considered,
the algorithm maintains two versions of the databases, each one implemented
by a hash map.  The ``current'' database contains the data of all signatures
which represent the current state, and the ``new'' database will include the
data of all new signatures that will be created by considering all current
signatures and the options for the current cell:  occupied (belonging to the
polyomino(es)) or empty (not belonging).
Figure~\ref{fig:boundary} shows a partially-built polyomino whose signature
(from top to bottom)\footnote{
   In Jensen's notation, signatures are specified bottom to top.
} is ``\texttt{23014024}''.

The keys to the database are numbers that represent the signatures, and the
mapping is from a signature to the data associated with it:  A statistics of
all partially-built polyominoes having the boundary that is represented by
that signature.
The data that correspond to a signature~$\sigma$ comprise the
\emph{counts} of partial polyominoes of all possible sizes, whose right
boundaries have the signature~$\sigma$.
We emphasize that the database stores only these statistics but does not
store all possible polyominoes.
Therefore, the main factor that dominates the memory consumption and the
running time of the algorithm is the \emph{number of signatures} and not
the number of (partially-built) polyominoes.  Both numbers are exponential
with~$n$, but as we detail below, the base of the former quantity is much
smaller than that of the latter quantity.
All other factors that influence consumed memory and running time are much
less significant.
This gives a huge advantage to the Conway-Jensen algorithm over that of
Redelmeier~\cite{redelmeier1981counting} since the latter produces all polyominoes one by one.

The basic step of the algorithm considers the next cell $t$ (refer again to
Figure~\ref{fig:boundary}), located just below the ``kink'' of the boundary.
In every such step, there are two options:  Either adding the cell~$t$ to
the polyomino (and hence enlarging its size by~1), or not adding it (hence,
no change in size).
Either option moves the kink below cell~$t$, therefore,
the signature changes from~$\sigma$ to some other signature~$\sigma'$.
The new signature~$\sigma`$ is fetched on the fly from a hard-coded
transfer matrix that considers the two boundary cells adjacent to the kink.
However, in a few cases, the transfer matrix does not imply~$\sigma`$
directly, but the entire current boundary, $\sigma$, needs to be scanned
at running time.
The algorithm then updates the contents of the database entry with
key~$\sigma'$, using the contents of the entry with key~$\sigma$ and
a set of rules that depend on whether or not the cells above and to the
left of~$t$ are occupied or free, and sometimes on other cells along the
boundary.  The full set of rules is provided in
Refs.~\cite{jensen2001enumerations,jensen2003counting}.
The signatures do not include an indication about the location of the kink.
This is not much of a concern as the new signature~$\sigma'$ is stored in a
new copy of the database.  In practice, in each step of considering the next
cell~$t$, the algorithm keeps two copies, old and new, of the database.  All
possible signatures are processed while considering~$t$, and all updates are
performed in the new copy of the database.  Before moving to the next cell,
the new copy is copied into the old copy.

During the course of the algorithm, it sums up the counts found in all
entries that correspond to valid (connected) polyominoes up to some
prescribed size and for all possible heights ($y$-spans).
(Note that whether or not partial polyominoes are connected is manifested by
the signatures of their right boundaries.)
An implementation detail concerns the ``timing'' for this operation: 
Updating the counts of polyominoes is performed when the boundary ``loses
contact'' with the polyominoes to the left of it, that is, when the
boundary~$\sigma'$ becomes completely empty.

\emph{Remarks}. \\
1.~The algorithm needs only be run for all heights
$1 \leq h \leq \ceil{n/2}$ since all polyominoes whose height
exceeds~$\ceil{n/2}$ are also detected with their
90$^\circ$-rotated versions.
Hence, the algorithm counts only polyominoes whose width is at least their
height, and doubles the counts of polyominoes whose width is strictly bigger
than their height. \\
2.~However, in order to avoid multiple counting of polyominoes whose height
is strictly smaller than $\ceil{n/2}$ (since they fit into more than one
box), the keys (signatures) are extended with two bits that indicate whether
polyominoes touch the floor and/or ceiling of the bounding box.\footnote{
   In Jensen's terminology~\cite[P.~868]{jensen2001enumerations}, the string
   encoding the boundary is a ``configuration,'' and together with these two
   bits, they make a ``signature.''
}
(This information is not reflected in the original signatures since the
touching point(s) can be to the left of the boundary instead of on it.)
Thus, polyominoes are counted only for entries whose two additional bits are
set.

Analysis of the algorithm~\cite{barequet2007complexity}
shows that the major factor in its running time and memory consumption is
the number of signatures it processes.
It turns out that the number of possible signatures of length~$h$ is roughly
proportional to $3^h$.  (This is proven by a bijection between signatures
and Motzkin paths.)  The amount of storage in each entry of the database,
and the time needed for accessing and processing each such entry are
proportional to low-degree polynomials in~$n$.
Hence, the running time of the algorithm is
roughly comparable (up to a small polynomial factor) to
$3^{n/2} \approx 1.73^n$, a quantity that is significantly smaller
than~$4.06^n$, which is roughly the number of polyominoes of size~$n$.
Pruning of signatures (discarding signatures of partial polyominoes that
cannot grow further and form valid polyominoes) is a very efficient
enhancement which Jensen introduced to the algorithm.
Without pruning, the taller the bounding box is, the more time-consuming
its processing is.  (This is implied immediately by the fact that the number
of signatures depends exponentially on the height of the box.)
It turns out that when signatures are pruned, the height of the box that is
the most computationally heavy becomes smaller than~$n/2$, and for higher
boxes, pruning is more effective and the running time for processing them
drops significantly.
Hence, at least for small values of~$n$, when pruning is used, the running
time behaves even better than~$3^{n/2}$.
As we describe below, we introduce a crucial improvement to the
Conway-Jensen algorithm, and thereby compute many more terms of the
sequence~$(A(n))$, extending the count by~14 new terms.


\section{Our Improvement}
\label{sec:our-improvement}

The Conway-Jensen algorithm essentially computes the number of $n$-cell
polyominoes bounded by axis-aligned boxes of various heights (originally
called ``widths'').
In a nutshell, our improvement is achieved by rotating the bounding boxes by
$45^\circ$ and scanning them along diagonals (of the original lattice).
(Alternatively, we rotate the \emph{lattice} by $45^\circ$ and scan it by
columns.)

For describing our modification, we have to define the
\emph{diagonal height} of a polyomino.
First, we need to assign sequential ordinal numbers to the diagonals of the
lattice.
Figure~\ref{fig:diag_orders}(a)
\begin{figure*}
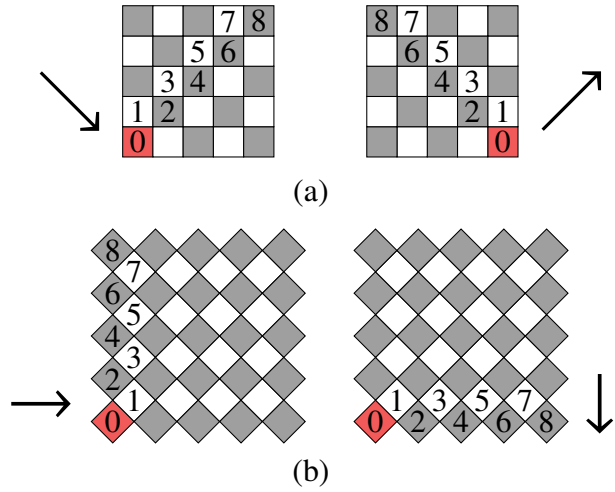

   \centering
   \begin{tabular}{cc}
      \raisebox{6mm}{\Huge $\searrow$} \drawpoly{diag_order1.txt} &
         \drawpoly[yscale=1,xscale=-1]{diag_order1.txt}
         \raisebox{6mm}{\Huge $\nearrow$} \\
      \multicolumn{2}{c}{(a)} \medskip \\
      \raisebox{3mm}{\Huge $\rightarrow$}
         \drawpoly[rotate=45]{diag_order2.txt} &
         \drawpoly[rotate=45,yscale=-1,xscale=1]{diag_order2.txt}
         \raisebox{3mm}{\Huge $\downarrow$} \\
      \multicolumn{2}{c}{(b)}
   \end{tabular}
   \caption{The numbering of diagonals (the red cell is the origin~(0,0)):
      (a) Left: Descending diagonals~($D_1$);
          Right: Ascending diagonals~($D_2$).
      (b) The same diagonals on the rotated lattice, becoming rows and
          columns, respectively.}
   \label{fig:diag_orders}
\end{figure*}
shows the numbering of descending and ascending diagonals.
Then, we need to assign to every cell coordinates that are based on this
numbering.  Formally, we associate with every cell~$c=(x,y)$ the two
diagonals it resides on using the functions $D_1(x,y) = x+y$ and
$D_2(x,y) = x-y$, that is, the coordinates of~$c$ in the rotated lattice 
are~$(D_1(x,y),D_2(x,y))$.
The two ``diagonal distances'' (horizontal and vertical) between two cells,
$(x_1,y_1)$ and $(x_2, y_2)$, are $\abs{D_1(x_1,y_1) - D_1(x_2,y_2)}$
and $\abs{D_2(x_1,y_1) - D_2(x_2,y_2)}$, respectively.

For ease of exposition, we will hereafter draw the lattice rotated
by~$45^\circ$ as in Figure~\ref{fig:diag_orders}(b).
In this orientation, each original diagonal becomes a row or a column,
and so the terms ``diagonal width'' and ``diagonal height'' of a polyomino
are defined as the maximum horizontal and vertical distance between two
cells in the rotated lattice, plus one.

Let us now define \emph{axis-aligned boxes} on the $45^\circ$-rotated
lattice.
A box of width~$w$ and height~$h$ contains all cells found in between and
including two columns within distance~$w$ and two rows within distance $h$.
In contrast with boxes on the regular (unrotated) lattice, the shape of such
a box may take several forms, depending on the parities of the columns and
rows of its boundary;  See Figure~\ref{fig:boxes} for a few examples.
\begin{figure}
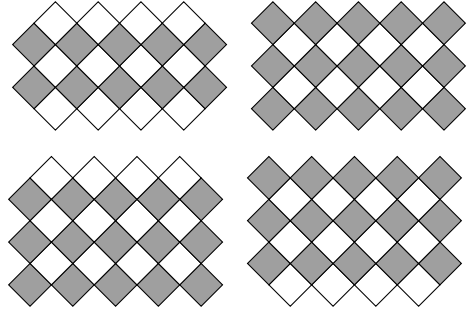

    \centering
    \drawpoly[rotate=45]{diag_rect1.txt}
    \drawpoly[rotate=45]{diag_rect2.txt}
    \\[3mm]
    \drawpoly[rotate=45]{diag_rect3.txt}
    \drawpoly[rotate=45]{diag_rect4.txt}
    \caption{A few types of boxes.  The two top boxes are of height~$5$ and
             width~$9$, however, they are are essentially different.
             The two bottom boxes are of height~$6$ and width~$9$;  these
             boxes are different as well but one can be obtained from the
             other by mirroring vertically.}
    \label{fig:boxes}
\end{figure}

Similarly to the original version of the algorithm, our version counts the
different polyominoes that span the entire box, \emph{i.e.}, touch its four
sides, for all possible sizes of axis-aligned boxes.
An alternative interpretation of a polyomino spanning a box is saying that
it has the same diagonal height and width (regular height and width on the
rotated lattice) as the box it spans.
Figure~\ref{fig:spanning}
\begin{figure}
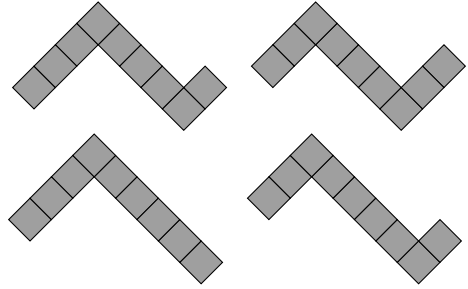

    \centering
    \drawpoly[rotate=45]{span1.txt}
       \drawpoly[rotate=45]{span2.txt} \\
    \drawpoly[rotate=45]{span3.txt}
       \drawpoly[rotate=45]{span4.txt}
    \caption{Polyominoes which span the respective boxes in
             Figure~\protect\ref{fig:boxes}.}
    \label{fig:spanning}
\end{figure}
shows a few polyominoes which span the respective boxes shown in Figure~\ref{fig:boxes}.

As in the original version of the algorithm, we simulate the process
of building polyominoes cell by cell, while keeping the counts of
partially-built polyominoes for every possible boundary signature.
Cells on the boundary are classified according to the components they
belong to in the same manner as in the original version of the algorithm.
An example of a partial polyomino and the respective signature is shown in Figure~\ref{fig:partial}.
\begin{figure}
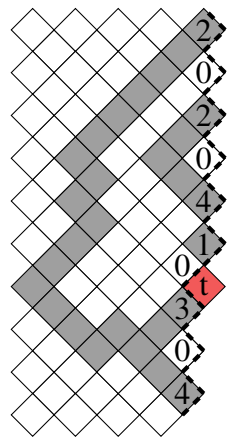

    \centering
    \drawpoly[rotate=45]{partial.txt}
    \caption{A sample partial polyomino~$P$, with signature ``2020410304'',
             built in the course of the algorithm.
             The signature of~$P$ is the sequence of numbers on its boundary.
             The current boundary is marked with a dashed line and the next
             cell to be considered, $t$, is colored in red.}
    \label{fig:partial}
\end{figure}
We say that the ``kink'' of the boundary is
in the transition of the boundary from a column in the rotated lattice
(diagonal in the original lattice) to the column immediately to the left of
it.  Figure~\ref{fig:partial} shows the kink wrapping the red cell~$t$ on
its left side.
When cell~$t$ is considered, the types of the cells on its top-left
and bottom left sides (denoted as `T' and `B', respectively),
as well as whether the next cell is occupied or empty, imply the
modified signature.  Such operations are stated as ``transition rules.''
Obviously, the transition rules in the modified algorithm are different from
those in the original algorithm.
The modified rules are shown in Table~\ref{tab:transition}.
\begin{table}
    \centering
    \caption{The modified transfer matrix (actions taken when processing the
             next cell~$t$, the cell under the kink).
             }
    \begin{tabular}{|c|r|c|c|c|c|}
    \hline
    T~\textbackslash~B &\multicolumn{1}{c|}{~~0} &1 &2 &3 &4 \\
    \hline
    0 & $00$~ $10$ & $01$ $24$ & $02$ $23$ & $03$ $33$ & $04$ $34$ \\
    \hline
    1 & add $10$ & --- $24$ & --- $23$ & --- $33$ & --- $34$ \\
    \hline
    2 & $\overline{00}$~ $20$ & $\overline{01}$ $23$ & $\overline{02}$
           $\widehat{23}$ & $02$ $23$ & $01$ $24$ \\
    \hline
    3 & $00$~ $30$ & $01$ $33$ & $02$ $\widehat{33}$ & $03$ $33$ & $04$
           $34$ \\
    \hline
    4 & $\overline{00}$~ $40$ & $\overline{01}$ $34$ & $\overline{02}$
           $33$ & $\overline{03}$ $\widehat{33}$ & $\overline{04}$
           $\widehat{34}$ \\
    \hline
    \end{tabular}
    \label{tab:transition}
\end{table}
In this table, the left column and the top row, respectively, specify the
types of the cell on the top-left (T) and on the bottom-left (B) sides of
the kink cell~$t$ (colored in red in Figure~\ref{fig:partial}).
Each entry corresponds to a pair of cells (T and B) around the kink, and it
contains two actions, the left (resp., right) to be performed when the
considered kink cell becomes empty (resp., occupied).
Each operation is made of two digits:  The types of the two cells around
the new kink cell.
However, like in the original version of the algorithm, for certain
configurations (T and B cells, and whether the considered cell becomes
empty or occupied), the operation is not local but rather requires the
scanning of the entire boundary and potentially updating the types of other
cells along it.
There are two types of such global actions, denoted by a bar or a hat above
the 2-digit action.  A bar denotes the case in which the first or last cell
of a connected component disappears from the boundary, and thus, another
cell needs to be marked as the first or last boundary cell of this
component.
The hat sign denotes the case in which two different components are united
into one component, hence, the signature needs to be adjusted accordingly.
(See Ref.~\cite{jensen2003counting} for the full details.)
The action ``add'' means that the statistics of the current signature
should be added to the total count if the respective polyominoes span the
entire height of the box.  (Recall the two bits of the signature that
indicate this.)


\section{Implementation Details and Differences}
\label{sec:implementation}

In this section, we describe implementation details and differences between
the two versions of the algorithm.

\subsection{Arithmetic}

While Jensen's implementation used modular arithmetic~\cite{knuth2014art}
in order to represent big numbers, we used in our implementations of both
versions of the algorithm the C++ type\, \verb+long int[3]+, that is,
192-bit integers which can hold numbers up to about $6.2 \times 10^{57}$.
This was possible (and much more efficient) since the only
needed operations were initialization by~0 and addition.

\subsection{Pruning}

Given a target maximum polyomino size, a key idea for the efficiency of the
Conway-Jensen algorithm is the pruning of signatures representing partial
polyominoes that cannot be completed into valid polyominoes.
Signatures of partially-built polyominoes are kept by the algorithm only if
they can eventually develop into valid polyominoes, which will be counted
only if they satisfy the following three requirement:
\begin{itemize}
\item[(a)]
   (connectivity)~They will have only one connected component;
\item[(b)]
   (height)~They will touch both the top and bottom of the bounding box;
   and
\item[(c)]
   (width)~They will be as wide as high.
\end{itemize}
The first requirement is for counting only valid (connected) polyominoes,
and the second and third requirements are needed for avoiding multiple
counting of polyominoes.  As we explain below, some intermediate signatures
represent polyominoes that cannot evolve into valid polyominoes, and hence,
these signatures can be discarded (the earlier the better) in order to save
running time later.

For any signature~$\sigma$, we measure how far it is from satisfying these
requirements by using three parameters:
\begin{enumerate}
\item $n_c(\sigma)$:  The minimum number of additional cells needed for
      connecting all components into one component;
\item $n_h(\sigma)$:  The minimum number of cells needed to make
      partially-built polyominoes, having~$\sigma$ as their boundary, touch
      both the top and bottom of the bounding box; and
\item $n_w(\sigma)$:  The minimum number of cells needed to make the width
      reach (or exceed) the height.
\end{enumerate}
Depending on the signature, each of these three parameters can be~0.
For brevity of exposition, we omit the dependencies on $\sigma$ when they
are clear from the context.
The first quantity, $n_c$ is demonstrated in Figure~\ref{fig:pruning}.
The other two quantities are analogous.
\begin{figure}
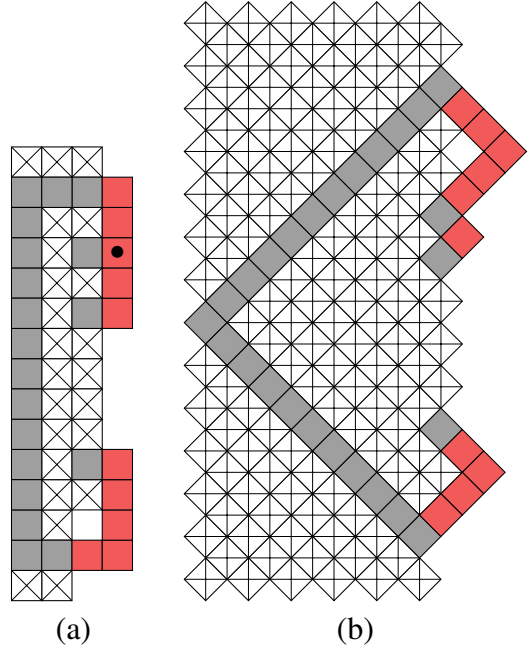

    \centering
    \begin{tabular}{cc}
       \drawpoly{inner_connection1.txt} &
          \drawpoly[rotate=-45]{inner_connection2.txt} \\
       (a) & (b)
    \end{tabular}
    \caption{An illustration of~$n_c$ in both lattices:
             Partial polyominoes and the smallest sets of cells (colored in
             red) needed to make them connected.}
    \label{fig:pruning}
\end{figure}

In the original version of the algorithm, where the lattice is in its regular
orientation, these quantities are ``almost independent'' in the following
sense.
For determining the value of~$n_c$, the connecting sequences are
(imaginarily) constructed along the column immediately to the right of the
current boundary (see Figure~\ref{fig:pruning}(a)).
This implies that they overlap in one column with
the cells needed for satisfying the width requirement, affecting~$n_w$.
This explains the term ``+1'' in the ``pruning formula'' specified below.
When we count polyominoes up to size~$n$, any signature that corresponds to
partial polyominoes of current size $n-(n_c+n_h+n_w)+1$ or more can be
discarded since completing the respective partial polyominoes into valid
(conneceted) poyominoes would exceed a total size of~$n$.

Note that setting a target size~$n$ is not necessary for running
the algorithm.  However, it enables the pruning mechanism which reduces
significantly the number of signatures (whose number is exponential
with~$n$), which is the main factor in both memory consumption and running
time.

Unlike the original version, these three parameters are very much
interrelated in the $45^\circ$-rotated lattice version of the algorithm.
First, cells added diagonally in order to touch the top and bottom of the
bounding box also contribute to the width.
Therefore, we need to add only~$\max{(n_h,n_w)}$ cells in order to make
polyominoes satisfy both requirements~(b) and~(c).
Second, when connecting components using ``L-shaped'' sequences
(see Figure~\ref{fig:pruning}(b)), these
sequences also contribute to the width.
Suppose then that we need~$k$ sequences of lengths
$(s_1,\ldots,s_k)$ for connecting all the components.  Each sequence of
length~$s_i$, when built as an L-shape (which is the most ``economical''
way in order to gain on the way more width), contributes at
most~$\ceil{s_i/2}$ to the width.
Let $s^* = \max_{1 \leq i \leq k}{s_i}$.
Then, at least $\max{\set{n_w, n_w+n_c-\ceil{s^*/2}, n_h+n_c}}$
additional cells are needed to satisfy all three requirements.
Note that the L-shaped connection may extend beyond the right side of the
bounding box, therefore the second term might be negative, and hence the
first term cannot be omitted.
Also note that the sequence of length~$s^*$ is not necessarily included
in the optimal (minimum length) combination of paths that makes the
polyomino connected, thus, using it would make the polyomino wider
than necessary.  This consideration is taken into account in the second
term of the maximization.

\vspace{2mm}  
\noindent
\emph{Remarks}.
\begin{itemize}
\item
While computing~$n_h$ and~$n_w$ is straightforward in both versions of the
algorithm, computing~$n_c$ requires finding a minimum-length set of sequences
connecting all components. In the original version of the algorithm, where
connecting sequences may overlap at their endpoints (see the cell marked by
a bullet in Figure~\ref{fig:pruning}(a)), this can be solved by using a
dynamic-programming procedure.
(We do not know whether or not this approach was used by
Jensen~\cite{jensen2003counting} since he he did not mention this
explicitly.)
In our version, where the connecting sequences cannot overlap, we initially
used this approach.  However, we later switched to a procedure for
computing a minimum spanning tree (MST) of a graph in which the nodes were
the connected components, the edges were the sequences connecting the
components, and the weights of the edges were the lengths of the sequences.
The latter approach proved to be in practice faster than the former
approach, at least for the small number of connected components we dealt
with.
Note that the MST approach would not work in the original version of the
algorithm since the overlaps of sequences make the problem
equivalent to the quadratic minimum-spanning-tree
problem~\cite{assad1992quadratic}, which is NP-Hard in general.
At any rate, since we are dealing with very small values of~$n$, we are
interested in actual rather than asymptotic running times.
Moreover, our measure of efficiency, by which we compare the two
versions of the algorithm (see Section~\ref{sec:comparison}), is the number
of pruned signatures, which is independent of how we implement the
computation of~$n_c$.
\item
In order to appreciate how much complicated it is to compute the number of
cells needed for making sure that the final polyominoes are connected, we
refer the reader to Knuth's documented code in his implementation of the
Conway-Jensen algorithm~\cite{knuth2001polynum}.
\end{itemize}

\subsection{Symmetries}
\label{sec:symmetries}

Another method for reducing the number of signatures whose data are stored
in the database is utilizing symmetries that show up in the course of the
algorithm.  We now detail how we treat symmetric signatures in the
45$^\circ$-rotated lattice.

The first type of symmetry arises when we consider different boxes.
When the height is even, we have pairs of boxes that are identical up
to mirroring vertically, \emph{e.g.}, the two lower boxes in
Figure~\ref{fig:boxes}.
This allows us to process only one of these boxes, and then double the
respective final count.

The second type of symmetry arises in self-symmetric boxes of odd height.
Consider ``kinkless'' signatures, that is, signatures created when we finish
processing a column.
If two such distinct signatures are the inverse of each other, then the
respective partial polyominoes can be completed into valid polyominoes in
the same number of ways.
This allows us to merge every such pair of signatures into a single one,
roughly halving the number of signatures before starting to process the
next column.
Note that mirroring signatures is not applicable to boxes of an even height since doing this would project the respective partial polyominoes into
another, mirrored, box of the same height. For example, consider any
signature in the lower left box in Figure~\ref{fig:boxes}.
The three cells making the signature are in rows~1, 3, and~5
(starting to count top-down from~0).
Flipping it vertically would map the signature cells into
rows~0, 2, and~4.  Hence, it cannot be part of this box.
Although this technique reduces the number of signatures only at the
beginning of processing a new column, we found out that processing a
non-negligible amount of additional cells was needed for the set of
signatures to grow back to the full set (with respect to
the same run without utilizing this symmetry).  In practice, this
reduced the number of signatures by a considerable amount of roughly 25\%.

This improvement is applicable also to the original version of the algorithm.
We did not find a mentioning of this in the respective
works~\cite{jensen2001enumerations,jensen2003counting} but we included it in
our implementations of both versions.

\subsection{Parallelization}

The parallelization scheme that we used was a generalization of the one
proposed in the original version of the algorithm.
Jensen suggested~\cite{jensen2003counting} that at the beginning of
processing each column, the signatures be split into buckets according
to the occupancy pattern of the \emph{lower} half of the signature, where
the ``occupancy pattern'' is the binary pattern of boundary cells when
considering only whether or not they are occupied, disregarding the
codes 1--4 of these cells, that is, ignoring the connectivity to the left
of the boundary.
Since placing an occupied cell in the top half of the boundary
would not change the occupancy pattern in the lower part (a nice
observation in Ref.~\cite{jensen2003counting}), and, thus, the new
signature would be in the same original bucket, such buckets could be
processed independently of each other and in parallel until we reach the
lower half of the signature.
At this stage, the signatures are redistributed into new buckets, this
time according to the occupancy pattern of the \emph{upper} half of the
pattern, with the same reasoning as above.

In the modified version of the algorithm, we added a parameter that
controlled the granularity of the split into different buckets of
signatures.  When the parameter was set to $k$, we split the signatures
into~$k$ chunks (instead of only two chunks), and redistributed the signatures
before processing each chunk of~$h/k$ cells of the boundary,
where~$h$ was as before the length of the signature (height of the bounding
box).\footnote{
   In cases where~$h$ was not divisible by~$k$, the size of the
   first~$(h \bmod k)$ chunks was rounded up and the size
   of the rest of the chunks was rounded down.
}
When signatures were redistributed, they were split into buckets according to
the occupancy pattern of the complementing part of the chunk, which was of
size~$(k-1)h/k$.
Obviously, setting $k=2$ would result in the original parallelization scheme
of Jensen.

Using values of~$k$ larger than~2 incurred a larger time spent on
redistributing the signatures between the new buckets.  However, this
modification allowed us to split the signatures into more buckets, which
was eventually beneficial for two reasons.
\begin{enumerate}
\item
First, more buckets allowed a better balancing of the workload between
processors.
Consider, for example, counting polyominoes of size $n=56$, for which
the most time-consuming box in our implementation was of height~15.
(According to Jensen, the most time-consuming box in his implementation
for this size was of height~22--24.  
As discussed in Section~\ref{sec:comparison}, the bottleneck of the
algorithm in our version was met in a smaller
box height.\footnote{
   We remind the reader that unlike in our version of the algorithm, the
   signature length in the original version was identical to the height of
   the bounding box.
})
When the signature length was~15 and using only $k=2$ buckets, the patterns
were split according to half of the cells, that is, 7 (or~8) cells, thus,
into~128 (or~256) buckets, say~128.
These buckets were in practice very far from being equal in size.
For example, there were many more signatures with an occupancy pattern
of~$1010101$ than patterns with occupancy pattern~$0000000$.
When we increased~$k$ to, say, 5, we now used an occupancy pattern of
size~12 (four chunks of size~3), resulting in 4096 buckets which allowed
much more flexibility in balancing the load between the different processors.
\item
Second, a major motivation for this modification was reducing memory
consumption, which was the main computational bottleneck when counting
polyominoes of large sizes.
Each processor processed a single bucket of signatures at a time, while
non-active buckets were stored in a compressed way on a hard drive
(external memory).
When we split the signatures into more buckets, each bucket was smaller,
and, hence, the total memory requirement for the active buckets was
reduced because we had less uncompressed database entries of cells at a
time.
\end{enumerate}


\section{Comparison}
\label{sec:comparison}

Some factors play in favor of our version, compared to the original version
of the algorithm, while other factors play the opposite role.  Empirically,
our version was more efficient for larger polyominoes, and we now attempt to
explain why.

\begin{enumerate}

\item \emph{Number of boxes}.
    Recall that for computing the number of polyominoes of size~$n$, we count
    polyominoes in boxes of a range of heights. In the original version of 
    the algorithm, heights are in the range $1,\dots,\ceil{n/2}$.
    The algorithm stops at height~$\ceil{n/2}$ since we count only
    polyominoes whose width is at least their height.
    In our version of the algorithm, we cannot apply this optimization.
    Although our version also counts only polyominoes whose width is at
    least their height (in terms of diagonal width and height), it cannot
    stop at boxes of height~$\ceil{n/2}$ since boxes of greater heights still
    contain polyominoes that are wider than their height and at the same
    time span the entire box.
    A simple example is the ``stick'' of length~$n$
    \mbox{(\hspace{-1mm}\drawpoly[scale=0.5]{stick.txt}\hspace{-1mm})}.
    Both its diagonal width and height are~$n$, and, therefore, its two
    orientations are counted by our version of the algorithm only within
    the box of height~$n$.
    Consequently, our version needs to run till boxes of height~$n$,
    while the Conway-Jensen version stops at height~$\ceil{n/2}$.
    In addition, recall that for each height, we have two types of boxes,
    and, thus, we have a total of~$2n$ boxes to iterate through, compared
    to~$\ceil{n/2}$ in the original version of the algorithm.

\item \emph{Signature length}.
    The parameter that affects the most the running time of the algorithm
    is the length of the signature.
    In the original version of the algorithm, each row has a cell
    participating in the signature, and, thus, the signature length is equal
    to the height of the box.
    In our version of the algorithm, the participating rows are
    interleaved, and, thus, only half of the rows are represented in the
    signature.
    Hence, the signature length is only half the height of the box.
    On the other hand, the maximum height of a box in our version is~$n$,
    compared to~$\ceil{n/2}$ in the original version of the algorithm.
    In conclusion, the maximum signature length is the same in both versions.

\item \emph{Pruning}.
    As discussed above, there are three quantities that affect the pruning of
    signatures in the course of the algorithm:
    The number of cells needed to make the components associated with a
    specific signature connected, the distances to the top and bottom of the
    box, and the number of cells needed to make the width of the polyominoes
    at least as its height.
    All three parameters, in both versions of the algorithm, are obviously
    bounded from above by the height of the bounding box.
    The difference between the two versions is that the height of the box in
    our version is at most~$n$, while it is at most~$\ceil{n/2}$ in the
    original.  However, this is compensated by the that fact that in
    our version, the signature length is half of the height of the box.
    For all three parameters, the range of possible values is larger in our
    version, hence, the pruning potential is larger as well.

\item
    \emph{Adjacent boundary cells}.
    In the original version of the algorithm, two successive occupied
    boundary cells (which are not around the kink) touch each other, and
    hence always belong to the same connected component.
    This is not necessarily the case in our version of the algorithm.
    This difference is in favor of the Conway-Jensen version since it
    reduces in their case the number of possible signatures.
    
\item
   Signature pruning is (at least empirically) much more effective on the
   $45^\circ$-rotated lattice than on the regular lattice.  We attribute
   this to the fact that more cells (about twice) are needed on the former
   lattice than on the latter one for closing a vertical gap:
   For closing a gap of~$k$ cells, $2k+1$ cells are needed on the rotated
   lattice, while only~$k+2$ cells are needed on the regular lattice.
   This difference between the two versions is shown in
   Figure~\ref{fig:pruning}.
   In other words, signatures that contain big vertical gaps (between cells
   that belong to different components) are pruned in our version of the
   algorithm much earlier than in the original version.
\end{enumerate}

Putting together all the above, we see that there are factors in favor
and against each version of the algorithm when performance is compared.
Nevertheless, when we implemented both versions with all improvements
mentioned so far, our version was significantly more efficient for counting
polyominoes of larger areas.
This was mainly because it processed much fewer signatures, which are the
main factor in the running time of both versions.
We believe that Difference~(5) above was the reason for this.
In practice, given a specific target polyomino size, pruning of signatures
became effective much earlier in our version than in the Conway-Jensen
version of the algorithm, which was manifested in an earlier peak (shorter
signature length) with respect to time spent on processing boxes of
different signature lengths.
(See Figure~\ref{fig:times} for an illustration of this phenomenon for
counting polyominoes of size~40.)
Since for the same area, the bulk of the work was performed in our version
on shorter signatures, and signature pruning became more effective as they
became longer (as shown in Table~\ref{tab:sig_count}), our version was able
to process longer signatures (\emph{i.e.}, higher bounding boxes), thus it
was able to count larger polyominoes. 

Table~\ref{tab:sig_count}
\begin{table*}
    \centering
    \caption{The total number of signatures processed (after pruning) while
             computing~$A(n)$.  In our implementation of both versions, our
             version of the algorithm processes fewer signatures than
             Jensen's original version for~$n \geq 23$.}
    \begin{tabular}{|c|r|r|l|c|r|r|l|}
        \hline
        \multicolumn{4}{|c|}{Jensen's version wins} &
           \multicolumn{4}{c|}{Our version wins} \\
        \hline
        $n$ & \multicolumn{1}{c|}{Jensen's} & \multicolumn{1}{c|}{Ours} &
           \multicolumn{1}{c|}{Ratio} &
        $n$ & \multicolumn{1}{c|}{Jensen's} & \multicolumn{1}{c|}{Ours} &
           \multicolumn{1}{c|}{Ratio} \\
        \hline
        3 & 14 & 19 & 1.357 &
           23 & 2,197,843 & 2,174,613 & 0.989 \\
        4 & 21 & 65 & 3.095 &
           24 & 3,427,165 & 3,225,791 & 0.941 \\
        5 & 115 & 160 & 1.391 &
           25 & 5,402,306 & 4,793,829 & 0.887 \\
        6 & 160 & 396 & 2.475 &
           26 & 8,400,423 & 7,072,343 & 0.842 \\
        7 & 525 & 858 & 1.634 &
           27 & 13,114,620 & 10,392,875 & 0.792 \\
        8 & 719 & 1,702 & 2.367 &
           28 & 20,303,600 & 15,271,610 & 0.752 \\
        9 & 1,956 & 3,181 & 1.626 &
           29 & 31,472,506 & 22,383,133 & 0.711 \\
        10 & 2,798 & 5,819 & 2.080 &
           30 & 48,562,989 & 32,599,124 & 0.671 \\
        11 & 6,080 & 10,060 & 1.655 &
           31 & 74,873,463 & 47,398,923 & 0.633 \\
        12 & 9,305 & 16,944 & 1.821 &
           32 & 114,986,657 & 68,927,927 & 0.599 \\
        13 & 17,753 & 28,114 & 1.584 &
           33 & 176,561,372 & 99,988,249 & 0.566 \\
        14 & 28,351 & 45,593 & 1.608 &
           34 & 270,108,311 & 144,385,132 & 0.535 \\
        15 & 48,871 & 72,357 & 1.481 &
           35 & 413,271,060 & 208,511,368 & 0.505 \\
        16 & 79,345 & 113,324 & 1.428 &
           36 & 629,893,445 & 301,121,651 & 0.478 \\
        17 & 131,663 & 177,658 & 1.349 &
           37 & 960,526,646 & 434,209,504 & 0.452 \\
        18 & 211,907 & 274,791 & 1.297 &
           38 & 1,459,702,899 & 624,050,985 & 0.428 \\
        19 & 344,120 & 418,826 & 1.217 &
           39 & 2,218,864,323 & 895,884,471 & 0.404 \\
        20 & 545,895 & 636,957 & 1.167 &
           40 & 3,362,894,943 & 1,282,445,284 & 0.381 \\
        \cline{5-8}
        21 & 878,922 & 966,832 & 1.100 &
          \multicolumn{4}{l}{\multirow{2}{*}{~\vdots}} \\
        22 & 1,380,912 & 1,451,606 & 1.051 \\
        \cline{1-4}
    \end{tabular}
    \label{tab:sig_count}
\end{table*}
and Figure~\ref{fig:sig_count}
\begin{figure*}
   \centering
      \includegraphics[scale=0.80]{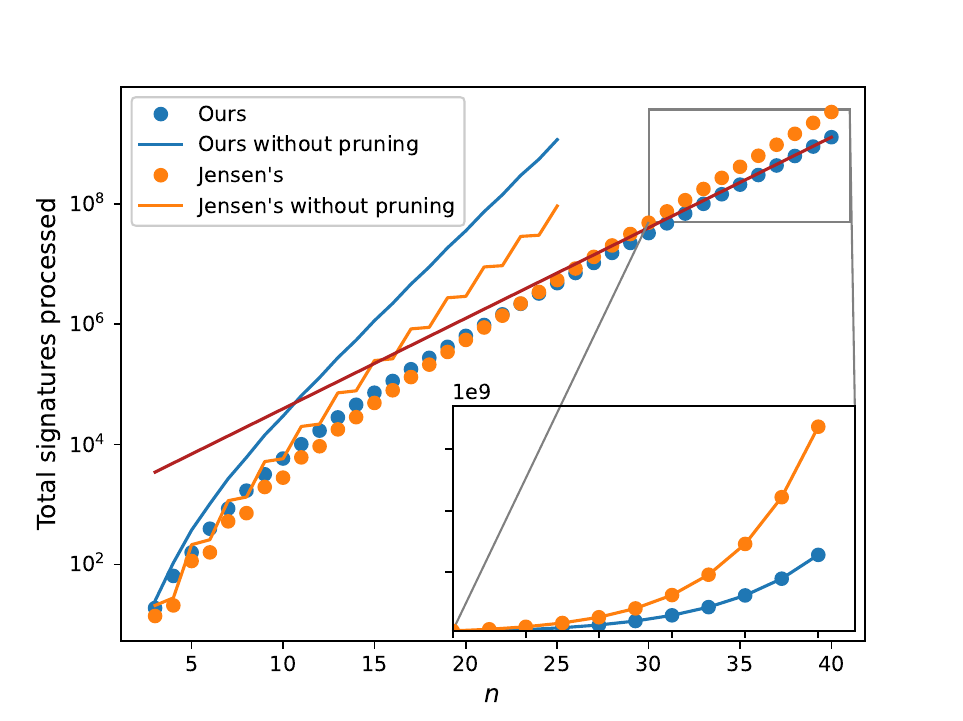}
      \includegraphics[scale=0.80]{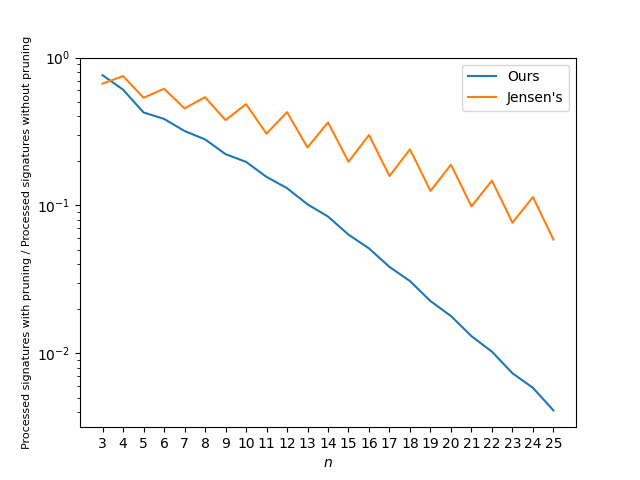}
   \caption{(Top) The total number of signature access operations
            while computing~$A(n)$ (in a logarithmic scale).
            The red line represents a growth rate of~$\sqrt{2}$.
            The growth rate of our version of the algorithm drops below
            $\sqrt{2}$ at $n=49$.
            The zoomed-in area is displayed in a regular scale.
            (Bottom) Ratios of signature pruning in the two versions.}
   \label{fig:sig_count}
\end{figure*}
compare the total number of signatures processed by the two versions of the
algorithm.\footnote{
   Just counting the number of signatures (after pruning) is misleading for
   two reasons.
   First, even if a signature is pruned, the algorithm still spends time on
   computing it.
   Second, some signatures are processed many times, which needs to be taken
   into account.

   Note that Jensen~\cite[P.~873, Fig.~2]{jensen2001enumerations} reports
   a different measure---the ``maximal numbers of configurations,'' which we
   interpret as the maximum number of configurations (not of signatures, see
   Section~\ref{sec:jc-algorithm}), taken over all individual considered
   cells (immediately under the kink) during the course of the algorithm.

   On top of all this, it is certainly possible that Jensen implemented
   pruning rules that are more efficient than those we implemented.
   However, as much as we can tell, we implemented all rules described in
   Refs.~\cite{jensen2001enumerations,jensen2003counting} and added more of
   our own.
   Anyway, in order to make a fair comparison between the two versions
   of the algorithm, we implemented the same pruning rules in both
   versions, and applied the same signature counting method in both
   implementations.
}
The symmetries improvements, discussed in Section~\ref{sec:symmetries}, were
implemented in both versions of the algorithm.
It can be seen that for small values of~$n$, the original version processes
fewer signatures than our version, however, starting at~$n=23$, the ratio
between the respective numbers becomes consistently more and more in favor of
our version.
Note that in the original version, the curve plotting the numbers of
signatures is ``jumpy'' since a new box appears only in odd heights, so that
in even sizes, the original number of signatures does not change much but
fewer signatures are pruned with respect to the previous size.

It is also worth mentioning that Jensen claimed that the number of
signatures \emph{after pruning}, when computing~$A(n)$, behaved
proportionally to~$\sqrt{2}^n$.
(It was shown later~\cite{barequet2007complexity} that the number of
signatures \emph{without pruning} was proportional to~$\sqrt{3}^n$.)
In our implementation, the growth rate of the number of signatures drops
below~$\sqrt{2}$ at $n=49$, and it then continues to drop for larger values
of~$n$.
This leads us to suspect that it is only a coincidence that the behavior of
this sequence is roughly like~$\sqrt{2}^n$ for a considerable amount of the
leading terms.

Another measure of comparison is the signature length in which the running
time reaches a peak (for a given value of~$n$), that is, which box requires
the longest time to compute.  Jensen reported that for computing~$A(56)$,
the boxes that required the longest running time were of heights in the
range~22--24.  In our version, when we computed~$A(56)$, the running time
reached a peak for boxes of height in the range~30--32, which translates
(as already explained) into signatures of size 15--16.
We believe that the reason for the peak being earlier (at smaller heights)
is that for higher boxes, the pruning was much more efficient because the
pruning conditions were more restrictive in the sense that given a current
size and a target polyomino size, more additional cells were needed in our
version in order to satisfy these conditions.
\begin{figure}
   \centering
   \includegraphics[scale=0.7]{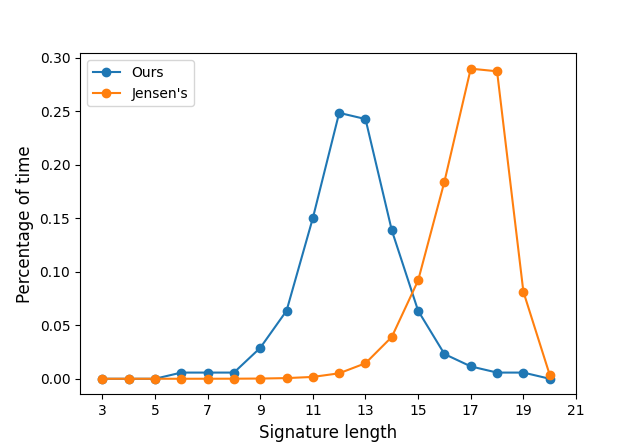}
   \caption{Percentage of time spent for each signature length when
            computing~$A(40)$.}
   \label{fig:times}
\end{figure}
Figure~\ref{fig:times} shows this behavior for the computation of~$A(40)$.
Recall that in our version of the algorithm, two consecutive heights have
the same signature length; moreover, different boxes have the same height.
Hence, for a comparison based on signature length,
we summed up times spent by our version of the algorithm on all boxes
that correspond to the same signature length.


\section{Results}
\label{sec:results}

We implemented the improved version of the algorithm in the \verb.C++.
programming language.  The computation of~$A(70)$ took about~$15,000$ CPU
hours (21 months) on a 32-CPU machine with 32GB of RAM and 250GB of disk
size used to store the data currently not being processed.
Table~\ref{tab:results}
\begin{table}
   \centering
   \caption{Counts of polyominoes.}
   \renewcommand{\arraystretch}{0.96}
   \begin{tabular}{|c|r||c|r|}
      \hline
      $n$ (size) & \multicolumn{1}{c||}{$A(n)$} & $n$ (size) &
         \multicolumn{1}{c|}{$A(n)$} \\
      \hline
      ~1 &              1 & 21 &                                         88,983,512,783 \\
      ~2 &              2 & 22 &                                        345,532,572,678 \\
      ~3 &              6 & 23 &                                      1,344,372,335,524 \\
      ~4 &             19 & 24 &                                      5,239,988,770,268 \\
      ~5 &             63 & 25 &                                     20,457,802,016,011 \\
      ~6 &            216 & 26 &                                     79,992,676,367,108 \\
      ~7 &            760 & 27 &                                    313,224,032,098,244 \\
      ~8 &          2,725 & 28 &                                  1,228,088,671,826,973 \\
      ~9 &          9,910 & 29 &                                  4,820,975,409,710,116 \\
      10 &         36,446 & 30 &                                 18,946,775,782,611,174 \\
      11 &        135,268 & 31 &                                 74,541,651,404,935,148 \\
      12 &        505,861 & 32 &                                293,560,133,910,477,776 \\
      13 &      1,903,890 & 33 &                              1,157,186,142,148,293,638 \\
      14 &      7,204,874 & 34 &                              4,565,553,929,115,769,162 \\
      15 &     27,394,666 & 35 &                             18,027,932,215,016,128,134 \\
      16 &    104,592,937 & 36 &                             71,242,712,815,411,950,635 \\
      17 &    400,795,844 & 37 &                            281,746,550,485,032,531,911 \\
      18 &  1,540,820,542 & 38 &                          1,115,021,869,572,604,692,100 \\
      19 &  5,940,738,676 & 39 &                          4,415,695,134,978,868,448,596 \\
      20 & 22,964,779,660 & 40 &                         17,498,111,172,838,312,982,542 \\
      \cline{1-3}
      41 & \multicolumn{3}{r|}{                                 69,381,900,728,932,743,048,483} \\
      42 & \multicolumn{3}{r|}{                                275,265,412,856,343,074,274,146} \\
      43 & \multicolumn{3}{r|}{                              1,092,687,308,874,612,006,972,082} \\
      44 & \multicolumn{3}{r|}{                              4,339,784,013,643,393,384,603,906} \\
      45 & \multicolumn{3}{r|}{                             17,244,800,728,846,724,289,191,074} \\
      46 & \multicolumn{3}{r|}{                             68,557,762,666,345,165,410,168,738} \\
      47 & \multicolumn{3}{r|}{                            272,680,844,424,943,840,614,538,634} \\
      48 & \multicolumn{3}{r|}{                          1,085,035,285,182,087,705,685,323,738} \\
      49 & \multicolumn{3}{r|}{                          4,319,331,509,344,565,487,555,270,660} \\
      50 & \multicolumn{3}{r|}{                         17,201,460,881,287,871,798,942,420,736} \\
      51 & \multicolumn{3}{r|}{                         68,530,413,174,845,561,618,160,604,928} \\
      52 & \multicolumn{3}{r|}{                        273,126,660,016,519,143,293,320,026,256} \\
      53 & \multicolumn{3}{r|}{                      1,088,933,685,559,350,300,820,095,990,030} \\
      54 & \multicolumn{3}{r|}{                      4,342,997,469,623,933,155,942,753,899,000} \\
      55 & \multicolumn{3}{r|}{                     17,326,987,021,737,904,384,935,434,351,490} \\
      56 & \multicolumn{3}{r|}{                     69,150,714,562,532,896,936,574,425,480,218} \\
      \hline
      \multicolumn{4}{|c|}{New Values\raisebox{3.5mm}{\mbox{}}} \\
      \hline
      57\raisebox{3mm}{\mbox{}} &
           \multicolumn{3}{r|}{           276,061,302,869,769,053,815,091,348,274,853} \\
      58 & \multicolumn{3}{r|}{         1,102,414,654,388,614,817,828,362,087,885,194} \\
      59 & \multicolumn{3}{r|}{         4,403,627,610,727,810,528,935,609,181,494,038} \\
      60 & \multicolumn{3}{r|}{        17,595,360,125,786,076,902,429,902,468,975,094} \\
      61 & \multicolumn{3}{r|}{        70,324,022,977,839,757,717,649,768,794,923,974} \\
      62 & \multicolumn{3}{r|}{       281,140,509,936,541,236,506,957,176,030,736,610} \\
      63 & \multicolumn{3}{r|}{     1,124,226,206,002,101,486,403,009,097,923,067,462} \\
      64 & \multicolumn{3}{r|}{     4,496,670,726,609,716,846,990,603,851,802,851,046} \\
      65 & \multicolumn{3}{r|}{    17,990,046,482,672,581,050,565,325,516,788,100,067} \\
      66 & \multicolumn{3}{r|}{    71,990,329,116,147,598,315,672,557,624,308,271,186} \\
      67 & \multicolumn{3}{r|}{   288,146,737,936,083,547,752,609,866,746,184,117,546} \\
      68 & \multicolumn{3}{r|}{ 1,153,580,914,959,613,979,832,549,218,358,894,513,290} \\
      69 & \multicolumn{3}{r|}{ 4,619,282,047,583,828,929,546,825,973,053,580,643,926} \\
      70 & \multicolumn{3}{r|}{18,500,792,645,885,711,270,652,890,811,942,343,400,814} \\
   \hline
   \end{tabular}
   \label{tab:results}
\end{table}
provides the counts of polyominoes computed by our program.
For sizes $1 \leq n \leq 56$, the values computed for~$A(n)$ agree with
previously published values~\cite{jensen2003counting}, while the values
for $57 \leq n \leq 70$ are new.

Let us attempt to estimate for which value of~$n$, our version of the
algorithm processes the same amount of signatures as does the
Conway-Jensen version for computing~$A(56)$ (the previously last known
element).
Knowing that for~$n=40$, the ratio of signatures processed by the two
versions of the algorithm is around~0.381 (see Table~\ref{tab:sig_count}),
and that the growth rates of numbers of signatures are almost steady (but
slightly monotone decreasing) around~1.5 and~$\sqrt{2}$, respectively, we
solve for~$s$ the equation $x \cdot 1.5^{16} = (0.381 \cdot x) (\sqrt{2})^s$,
and find that~$s \approx 21.5$.  Since the gap between the two sequences has
a slow tendency of growing, we round~$s$ up to~22, or even~23, and estimate
that our version of the algorithm processes roughly the same amount of
signatures at~$n=40+23=63$ as the previous version does for~$n=56$.
Hence, we attribute, out of the~14 new terms of~$A(n)$, about seven terms to
our improvement, and the other seven terms to other factors, such as having
better hardware, different implementations, and more running time available.

\mbox{} \\
\textbf{Remarks}
\begin{enumerate}
\item
In 1999, Madras~\cite{madras1999pattern} showed the existence of
the growth rate of polyominoes, the limit
$\lambda := \lim_{n \to \infty} (A(n+1)/A(n))$, and the currently
best-known lower bound on~$\lambda$ is~4.00253~\cite{barequet2016lambda}.
It is known \emph{empirically}, but has never been shown analytically, that
the sequence $(A(n+1)/A(n))$ is monotone increasing.  Had this been proven
rigorously, we would have obtained
$A(70)/A(69) > 4.00512$ as an improved lower bound on~$\lambda$.
\item
In a private communication, Iwan Jensen applied two numerical methods and
used the~14 new values of~$A(n)$ for obtaining~4.06256912(2) and
4.06256903(5) as improved estimates of~$\lambda$.
\item
G\"{u}nter Rote drew our attention to the fact that in a lecture in~2014,
D.E.~Knuth~\cite{knuth2014problems} had used diagonals on the square lattice
for deriving generating functions for the counts of polyominoes in given
bounding boxes, and provided enumeration of polyominoes of area~47 (rounded
to units of~$10^{22}$) by their bounding boxes.
About~14 years earlier, Knuth presented in a lecture\footnote{
   See the YouTube video
   \texttt{https://www.youtube.com/watch?v=df\_7hydjaT0}~.
} a similar method for building polyominoes in diagonals, where the idea was
to add an entire diagonal layer at a time.  This was much less efficient than
adding layers cell-by-cell as in the transfer-matrix algorithm.
\end{enumerate}


\section{Conclusion}
\label{sec:conclusion}

In this paper, we suggest a substantial improvement to the Conway-Jensen
algorithm for counting polyominoes.  Using a program that implemented this
improvement, we were able to extend the count of polyominoes from~56 terms
to the first~70 terms of the sequence~$(A(n))$.

Future research directions include introducing more programming tricks which
will allow us to compute more values of~$A(n)$ and generalizing the method
to three dimensions.


\section*{Acknowledgements}

The authors would like to thank G\"{u}nter Rote for helpful comments on
a preliminary version of this paper, and Iwan Jensen for
helpful discussions about the contents of this paper.
Special thanks also go to an anonymous referee which provided an excellent
critique of the paper which improved it tremendously.


\bibliographystyle{abbrv}
\bibliography{references}        

\end{document}